\documentclass[twocolumn,showpacs,amsmath,amssymb,superscriptaddress,pra]{revtex4}

\usepackage{graphicx}
\usepackage{amsmath,amssymb,amsthm}

\renewcommand{\vec}[1]{\boldsymbol{#1}}
\newcommand{\beq}{\begin{equation}}
\newcommand{\eeq}{\end{equation}}
\newcommand{\beqa}{\begin{eqnarray}}
\newcommand{\eeqa}{\end{eqnarray}}
\newcommand{\e}{\mathrm{e}}
\newcommand{\w}{\omega}

\newcommand{\ket}[1]{\left| #1 \right\rangle}

\newcommand{\der}[2]{\frac{\mathrm{d}#1}{\mathrm{d}#2}}

\newcommand{\ketbrax}[2]{\left|#1\right\rangle_x\hskip-1mm\left\langle #2\right|_x}

\newcommand{\ketx}[1]{\left| #1 \right\rangle_x}
\newcommand{\ketz}[1]{\left| #1 \right\rangle_z}

\begin{document}

\title{Long-lived spin entanglement induced by a spatially correlated thermal bath}
\author{D. P. S. McCutcheon} \email{dara.mccutcheon@ucl.ac.uk}
\affiliation{Department of Physics and Astronomy, University College London, Gower Street, London WC1E 6BT, United Kingdom}
\affiliation{London Centre for Nanotechnology, University College London}
\author{A. Nazir}
\affiliation{Department of Physics and Astronomy, University College London, Gower Street, London WC1E 6BT, United Kingdom}
\author{S. Bose}
\affiliation{Department of Physics and Astronomy, University College London, Gower Street, London WC1E 6BT, United Kingdom}
\author{A. J. Fisher} 
\affiliation{Department of Physics and Astronomy, University College London, Gower Street, London WC1E 6BT, United Kingdom}
\affiliation{London Centre for Nanotechnology, University College London}

\date{\today}

\begin{abstract}
 
We investigate how two spatially separated qubits 
coupled to a common heat bath can be entangled
by purely dissipative dynamics. We identify a new dynamical timescale associated with the lifetime of the 
dissipatively-generated entanglement and show that it can be much longer than either the typical single-qubit 
decoherence time, or the timescale on which a direct exchange interaction can entangle the qubits. We give an 
approximate analytic expression for the long-time 
evolution of the qubit concurrence and propose an ion trap scheme in which such dynamics should be observable.

\end{abstract}


\maketitle

\section{Introduction}
 
Entanglement is the hallmark of correlations in quantum theory and has  
come to be seen as a precious resource essential for many quantum information processing 
protocols~\cite{n+c}. An entangled state can show correlations stronger than those 
allowed classically, however, these correlations are 
often extremely fragile. Interactions with the environment surrounding any quantum 
system tend to cause a rapid 
loss of quantum coherence \cite{b+p}, 
generally leading to entanglement within the system being destroyed 
on a short timescale~\cite{Yu02}. 
Various techniques have thus 
been developed to protect entangled states from their surroundings, 
such as constructing decoherence free subspaces~\cite{lidar98, zanardi97}, dynamical decoupling~\cite{gordon06, fanchini07}, and exploiting the 
quantum Zeno effect~\cite{Maniscalco08}.

As well as avoiding decoherence,  
entanglement must also be generated.
This can be achieved by a number of means; for example, by harnessing intrinsic system couplings 
\cite{kraus01}, through projective measurements \cite{barrett05}, or via a quantum bus \cite{blais04}. Recently, it was shown that entanglement between a 
pair of two-level systems (qubits) can in fact be generated by the same processes that are usually considered to be detrimental, if the qubits are allowed to 
interact with a common bath~\cite{braun02,benatti03}. 
This offers a potential way to explore the interplay between coherent and incoherent multi-qubit dynamics  
with a significantly reduced level of external system control.

In general, immersing a pair of otherwise non-interacting spins in a common heat bath will give rise to two terms in the 
subsequent master equation: a unitary, Hamiltonian-like Lamb-shift term~\cite{b+p, solenov07} leading to coherent spin evolution, 
and a dissipative term~\cite{b+p,benatti03}, both of which may generate spin entanglement~\cite{braun02,benatti03,solenov07,an07,choi07}. 
While it is known that in the idealised case of unseparated qubits entanglement induced by the {\it dissipative} term can persist 
indefinitely~\cite{benatti05}, surprisingly little attention has been paid to the dynamics of its generation and 
decay in the more realistic setting of {\it finite} inter-qubit separation. 
In this experimentally relevant case, it is not clear on what timescale the generated entanglement persists, 
or even whether its observation is feasible at all. It has been shown that the entangling capability of the Lamb-shift 
is highly sensitive to the inter-qubit separation~\cite{solenov07}, with dissipative processes destroying any generated 
entanglement more rapidly as separation increases. Furthermore, it has recently been shown 
that the level of entanglement generated between two harmonic oscillators suffers from a similar critical dependence on oscillator separation~\cite{zell09}. 
Hence, a comparison of the timescales associated with conventional 
decoherence dynamics to those for dissipatively-induced spin-entanglement generation and decay is needed.

In this article, we address the above issues  
by studying the dynamics of bath-induced entanglement in the context of the two-spin-boson model, which has wide applications in the solid state 
and elsewhere \cite{leggett87, dube98}. We show that for small but finite spin separation, the timescale on which dissipatively-induced spin entanglement survives 
can be far larger than the corresponding single spin decoherence time. It also exceeds the timescale on which entanglement induced by either 
a {\it direct exchange interaction} or the Lamb-shift persists. In particular, we obtain an approximate analytic expression for the long-time 
dynamics of the two-spin concurrence and from this determine its survival time.  
We suggest an ion trap realisation of our model and 
demonstrate 
that observation of the generated entanglement should indeed be feasible.

\section{Master Equation}

We consider two spatially separated, identical, non-interacting spin qubits, 
each subject to a static field of strength $\Delta/2$ in 
the $x$-direction, and coupled to a common bath of harmonic oscillators. 
The Hamiltonian is given by
\beqa
H&=&H_S+H_I+H_B\nonumber\\
&=&-\frac{\Delta}{2}\sum_{n=1}^{2}\sigma_x^{n}+\sum_{n=1}^2\sigma_z^n\otimes B_n+\sum_k \w_k b_k^{\dagger}b_k,
\eeqa
with $\hbar=1$. Here, $\sigma_i^n$ is a Pauli operator acting on the $n$th qubit ($n=1,2$; $i=x,y,z$), $B_n$ is the bath 
operator coupling qubit $n$ to the bath, while $\w_k$ is the angular frequency, and $b_k^{\dagger}$ ($b_k$) the creation 
(annihilation) operator of the bath mode of wave vector $k$. 

To investigate the dynamics of the reduced two-spin density operator $\rho$
we assume that the qubit-bath coupling is weak compared to  
$\Delta$ and follow the standard Born-Markov and rotating-wave approximation approach~\cite{b+p}. 
This relies on a perturbative expansion in the system-bath coupling strength, and the assumption that the bath instantly re-thermalises after any interaction with the qubits. 
Setting $\langle H_I\rangle_B=0$, we obtain a Schr\"{o}dinger-picture master equation of the usual form
\begin{equation}
\der{\rho(t)}{t}=-i[H_S+H_{LS},\rho(t)]+\mathcal{D}(\rho(t)),
\label{eqrhodot}
\end{equation}
valid to second order in $H_I$. Here, the Lamb-shift provides a Hamiltonian-like 
contribution and is of the form 
\beq
H_{LS}=A(\sigma_x^1+\sigma_x^2)+B(\sigma_z^1\sigma_z^2+\sigma_y^1\sigma_y^2),
\eeq
where we have omitted a constant. The precise expressions for A and B depend on the details of the qubit-bath interactions and will be given in 
section~\ref{secbathcorrelationfunctions}. For now we will comment on their likely qualitative effects. 
The first term of $H_{LS}$ simply renormalises the static field strength due to the presence of the bath modes. 
The entangling capability of the second term deserves some attention since it represents an induced interaction between the two qubits, 
as has been explored in detail in Ref.~\cite{solenov07}. 
However, we will show that in general this entanglement decays at a rate far quicker than that generated by the dissipator, 
here given by 
\begin{align}
\mathcal{D}(\rho(t))=&\sum_{\w}\sum_{n,m=1}^2\gamma_{nm}(\w)\bigl( A_m(\w)\rho(t)A_n^{\dagger}(\w)\nonumber\\
&-\frac{1}{2} \{ A_n^{\dagger}(\w)A_m(\w),\rho(t)\} \bigr),
\end{align}
with frequency summation  
over the eigenvalue differences of $H_S$ ($\w=\pm\Delta$) and corresponding eigenoperators $A_n(\pm\Delta)=(1/2)(\sigma_z^n \mp i\sigma_y^n)$.

\section{Bath Correlation Functions}
\label{secbathcorrelationfunctions} 

The key quantities in the present discussion are the 
Fourier transforms of the bath correlation functions
$\gamma_{nm}(\w)=\int_{-\infty}^{\infty}\mathrm{d}s\e^{i \w s}\langle B^{\dagger}_n(s)B_m\rangle$,
for which we need a 
specific form for  
$B_n$.
As usual in the spin-boson model, we consider linear coupling between the qubits and the coordinate of each bath mode~\cite{leggett87, dube98, mahan}, 
such that $B_n(s)=\sum_k(g_k^nb_k^{\dagger}\e^{i\w_k s}+g_k^{n*}b_k\e^{-i \w_k s})$, with coupling constants $g_k^n$.  Note that with this form of 
$B_n$ our assumption of $\langle H_I\rangle_B=0$ leading to Eq.~(\ref{eqrhodot}) is justified.
An important aspect of this work is that the qubits have an explicit spatial separation and thus $g_k^1\neq g_k^2$. 
To make this evident, we consider our spins to be separated by a distance $d$ along the $z$-axis such that 
$g_k^1=g_k\e^{i d k \mathrm{cos\theta}/2}$ and $g_k^2=g_k\e^{-i d k \mathrm{cos\theta}/2}$, where $\theta$ is the 
polar angle measured against the $z$-axis in $k$-space, and $|g_k^1|=|g_k^2|=g_k$. 
Taking the bath to be in thermal equilibrium, we  
find
\beqa
\gamma_{12}(\w)&{}={}&\int_{-\infty}^{\infty}\mathrm{d}s\e^{i \w s}\sum_k|g_k|^2\big(N(\w_k)\e^{i\w_k s}\e^{ikd\cos{\theta}}\nonumber\\
&&\:{+}(N(\w_k)+1)\e^{-i \w_k s}\e^{-ikd\cos{\theta}}\big), 
\eeqa
where $N(\w_k)=[\mathrm{exp}(\w_k/k_B T_B)-1]^{-1}$ is the thermal occupation of mode $k$, $k_B$ is Boltzmann's constant, 
$T_B$ the temperature of the bath, $\gamma_{12}(\w)=\gamma_{21}(\w)$ (once the summation is performed), and $\gamma_{nn}(\w)$ is obtained by setting $d=0$.

Defining the bath spectral density to be $J(\w)=\sum_k|g_k|^2\delta(\w-\w_k)$~\cite{leggett87}, and an inverse dispersion relation (assumed isotropic) 
$k=\kappa(\w)$, we take the continuum limit of the summation over $k$ above to find 
\beq
\gamma_{12}(\w)=f(\kappa(\w)d)\gamma_{11}(\w),
\eeq
with $\gamma_{11}(+\Delta)=\gamma_{22}(+\Delta)=(N(\Delta)+1)\gamma_0$ and 
$\gamma_{11}(-\Delta)=\gamma_{22}(-\Delta)=N(\Delta)\gamma_0$, where 
$\gamma_0=2\pi J(\Delta)$ is the single-spin decoherence rate at zero temperature. 
Here, $f(x)$ describes the bath's spatial correlations and is determined by 
its dimensionality $D (=1,2,3)$. For $D=1$ 
we have $f(x)=\mathrm{cos}(x)$; for $D=2$, $f(x)=J_0(x)$, where $J_0$ is a Bessel function of the first kind; 
and for $D=3$, $f(x)={\rm sinc}(x)$. We thus write
\beq
\gamma_{12}(\pm \Delta)=(1-\delta)\gamma_{11}(\pm \Delta),
\eeq
where $1-\delta$ captures the degree of correlation between the baths seen by each qubit, becoming unity at $d=\delta=0$ (completely correlated) and, for 
$D>1$, zero as $d\rightarrow\infty$ ($\delta\rightarrow1$, completely independent). When $D=1$, the Markovian assumption constrains 
$\gamma_{12}$ to be periodic with respect to $d$. However, we will concentrate here on the limit where $d$ is small enough such that 
$\delta \approx (\kappa(\Delta)d)^2/2D$, for all $D$.

The strength of the two terms in the Lamb-shift Hamiltonian are given by combinations of the Hilbert transforms of the 
bath correlation functions. They are found to be
\beq
A=2\int_{0}^{\infty}J(\w)\mathrm{coth}\bigl(\w/(2 k_B T_B)\bigr)\left(\frac{\Delta}{\Delta^2-\w^2}\right)\mathrm{d}\w
\eeq
and
\beq
B=\int_0^{\infty}J(\w)f(\kappa(\w)d)\left(\frac{\w}{\Delta^2-\w^2}\right)\mathrm{d}\w
\eeq
where principal values are assumed. Note that for system-bath coupling in 2 or 3 dimensions $B\rightarrow0$ as the qubit separation is increased 
to infinity, expressing the fact that an uncorrelated bath can not give rise to any coherent coupling between the qubits. 
In contrast, $A$ contains no distance dependence since it represents a renormalisation of the single-qubit energy levels in each spin, independent of any bath correlations.

The relative strength of the coherent terms in the evolution, A and B, compared to 
the strength of the dissipative terms, given by $\gamma_{11}$ and $\gamma_{12}$, dictates whether the bath is capable 
of generating entanglement through the Lamb-shift~\cite{solenov07}. Evaluation of the relevant integrals involved necessitates a specific form 
of the bath spectral density. In this article we shall focus on the entanglement generated through dissipative processes and 
as such we may leave $A$ and $B$ unevaluated. In fact, we shall show in Seciton~\ref{sec:lamb} that the dissipatively induced entanglement can persist for 
times far larger than entanglement generated through the Lamb-shift, regardless of its strength.

\section{State Dynamics}

Rather than working directly with the reduced density operator 
$\rho$, it is more instructive to work in terms of a $16$ dimensional vector $\vec{\alpha}$, 
which is a generalisation of the Bloch vector for two-qubit states. 
It is constructed by flattening the matrix whose elements $\alpha_{ij}$ satisfy
\beq
\rho=\frac{1}{4}\sum_{i,j=0}^3\alpha_{ij}\sigma_i^1\otimes\sigma_j^2,
\label{eqnrhofromalpha}
\eeq
where $\sigma_0^1=\sigma_0^2=I$. The traceless property of the Pauli matrices ensures that 
$\alpha_{ij}=\langle\sigma_i^1\sigma_j^2\rangle$, and conservation of probability demands that $\alpha_{00}=1$. To describe the evolution of our system, 
we consider the eigensystem of the Liouvillian super-operator $\mathcal{L}$
defined (unconventionally) by $\dot{\vec{\alpha}}(t)=\mathcal{L}\vec{\alpha}(t)$, where the linearity of the 
transformation between $\rho$ and $\vec{\alpha}$ ensures that the dynamics generated by $\mathcal{L}$ 
is equivalent to that of Eq.~(\ref{eqrhodot}). Clearly, a state $\vec{\alpha}$ initially in an eigenstate 
of $\mathcal{L}$, say $\vec{\alpha}_l$ (with eigenvalue $\lambda_l$), will evolve 
according to $\vec{\alpha}(t)=\vec{\alpha}_l\e^{\lambda_l t}$. Hence, a general state  
evolves such that 
\beq
\vec{\alpha}(t)=\sum_{l=0}^{15} a_l\vec{\alpha}_l\e^{\lambda_l t},
\label{eqngenalpha}
\eeq
where the coefficients $a_l$ are determined from the initial conditions, and the sum runs over all eigenstates of $\mathcal{L}$.

We are primarily interested in the long-time dynamics of our system. Hence, it makes sense to search for eigenvalues of $\mathcal{L}$ with small (or zero) 
real parts since, assuming these parts are all negative, the corresponding eigenvectors will contribute towards the total state Eq.~(\ref{eqngenalpha}) 
on the largest timescale. 
We evaluate the full eigensystem of $\mathcal{L}$ analytically, 
though this leads to combersome expressions which we shall not give here. However, we can identify a number of important features for our subsequent analysis. For any non-zero qubit separation ($\delta\neq0$) there exists a single eigenvector, $\vec{\alpha}_0$, with zero eigenvalue, and all other eigenvalues have negative real parts. We therefore associate $\vec{\alpha}_0$ with the thermal 
state since it is that to which all states tend towards as $t\rightarrow \infty$. Of the remaining 15, there is a single eigenvalue, $\lambda_1$ (with
corresponding eigenvector $\vec{\alpha}_1$), which vanishes as $\delta\rightarrow0$ at all temperatures. Expanding the exact expression to first order in $\delta$ and second order in $N(\Delta)$ gives the simple form 
\beq
\lambda_1=-(1+3 N(\Delta))\delta\gamma_0.
\eeq 
It is also possible to show graphically that $\lambda_1$ varies approximately linearly with $N(\Delta)$ at all temperatures. 
For reasons that should become clear, we shall refer to the eigenvector corresponding to the null eigenvalue, $\vec{\alpha}_0$, and the eigenvector corresponding 
to the vanishing eigenvalue, $\vec{\alpha}_1$, as our eigenvectors of interest. The eigenvector corresponding to the thermal state, $\vec{\alpha}_0$, 
is expressible solely in terms of $R=(1+2N(\Delta))^{-1}=\mathrm{tanh}(\Delta/2 k_B T_B)$, and is given by
\begin{equation}\label{alpha0}
\vec{\alpha}_0:\{\alpha_{00},\alpha_{01},\alpha_{11},\alpha_{22}\}=\{1,R,R^2,0\},
\end{equation}
where the notation implies that the eigenvector has the elements specified, and that $\alpha_{22}=\alpha_{33}$, $\alpha_{01}=\alpha_{10}$, with all other elements being zero.
From a numerical analysis of $\vec{\alpha}_1$ we find that for $\delta \ll 1$, to a very good approximation $\vec{\alpha}_1$ can also be 
expressed just in terms of $R$, with corrections being of the order of $\delta$: 
\begin{equation}
\vec{\alpha}_1:\{\alpha_{00},\alpha_{01},\alpha_{11},\alpha_{22}\}\approx\{0,R,1+R^2,1\},
\label{alpha0alpha1}
\end{equation}
where the notation is the same as in Eq.~(\ref{alpha0}). 

There are two further notable eigenvalues, $\lambda_2$ and $\lambda_3=\lambda_2^{*}$. Once again, expanding the exact eigenvalues to first order in $\delta$, and to first order in $R$ about $R=1$, gives the expression
\beq
\lambda_2=\lambda_3^*=-(1/2)\gamma_0(1-R+2\delta -\delta R)-i\Delta.
\eeq
Note that these eigenvalues have vanishing real parts only in the limit that both the separation and temperature go to zero 
($\delta\rightarrow0$ and $R\rightarrow 1$). In either the zero temperature limit ($R\rightarrow1$) or the zero separation limit 
($\delta\rightarrow 0$) the corresponding eigenvectors, $\vec{\alpha}_2$ and $\vec{\alpha}_3=\vec{\alpha}_2^{*}$, are given by
\beq
\vec{\alpha}_2:\{\alpha_{02},\alpha_{03},\alpha_{12},\alpha_{13}\}=\{i,1,i,1\},
\eeq
where once again the notation implies the eigenvector has the elements specified but this time with 
$\alpha_{20}=-\alpha_{02}$, $\alpha_{30}=-\alpha_{03}$, $\alpha_{21}=-\alpha_{12}$ and $\alpha_{31}=-\alpha_{13}$, 
and all other elements being zero.

This leaves us with 12 eigenvalues to consider. Plotting their real parts as a function of $R$ it becomes clear 
that they are all $\sim-\gamma_0/R$, for all values of $\delta$. As such, the corresponding 
eigenvectors contribute towards the total state significantly only for times $t<\gamma_0^{-1}$ regardless of the temperature. These eigenvalues and eigenvectors shall 
be referred to as those with $l>3$. 

With the relative size of the real parts of the various eigenvalues in mind, we see that for small enough $\delta$, at 
times sufficiently greater than $\gamma_0^{-1}$, a general state can be approximated by
\begin{equation}
\vec{\alpha}(t)=\vec{\alpha}_0+a_1\vec{\alpha}_1\e^{\lambda_1 t}+a_2(\vec{\alpha}_2\e^{\lambda_2 t}\pm\vec{\alpha}_2^{*}\e^{\lambda_2^* t}),
\label{eqgenevolt}
\end{equation}
where we have normalised $\vec{\alpha}_0$, set $a_0=1$, and set $a_3=\pm a_2$ to ensure positivity of the corresponding density operator. 
The coefficient $a_1$ is found, by setting $t=0$
in Eq.~(\ref{eqngenalpha}), to be $a_1=(\Lambda-R^2)/(3+R^2)$, where 
$\Lambda=\langle \vec{\sigma}^1 \cdot \vec{\sigma}^2\rangle=\alpha_{11}+\alpha_{22}+\alpha_{33}$. 
For separable pure states $\Lambda$ represents a scalar product of single qubit Bloch vectors. Positivity of the corresponding density operator 
limits the range of $\Lambda$ to $-3\leq\Lambda\leq+1$.

To gain insight into the general features of a state described by Eq.~(\ref{eqgenevolt}), it is useful to write down the corresponding density operator 
in the zero temperature and separation limit, in which there are no decoherent processes due to the real parts 
of the relevant eigenvalues vanishing. Using Eq.~(\ref{eqnrhofromalpha}) we find
\beq
\begin{split}
\rho(t)=&(1+a_1)\ketbrax{\uparrow\uparrow}{\uparrow\uparrow}-a_1\ketbrax{\Psi^-}{\Psi^-}\\
&+a_2\sqrt{2}\Bigl(\e^{-i\Delta t}\ketbrax{\Psi^-}{\uparrow\uparrow}\pm \e^{i \Delta t}\ketbrax{\uparrow\uparrow}{\Psi^-}\Bigr),
\label{eqgenevolt2}
\end{split}
\eeq
where $\ketx{\Psi^-}=(1/\sqrt{2})(\ketx{\uparrow\downarrow}-\ketx{\downarrow\uparrow})=(1/\sqrt{2})(\ketz{\uparrow\downarrow}-\ketz{\downarrow\uparrow})$ 
is the Bell singlet, and 
$\ketx{\uparrow\uparrow}=(1/2)(\ketz{\uparrow\uparrow}+\ketz{\downarrow\downarrow}+\ketz{\uparrow\downarrow}+\ketz{\downarrow\uparrow})$. The 
subscripts $x$ and $z$ refer to the relevant basis, with our qubits defined with respect to $z$.
Written in this way, we can see that our state is a coherent mixture of the singlet 
and the state $\ketx{\uparrow\uparrow}$. At zero temperature, the state $\ketx{\uparrow\uparrow}$ is the ground state since it minimises the energy 
associated with the the system Hamiltonian and energy can not be absorbed from the environment. Also, at zero separation, the 
Bell singlet is stable since it is the totally anti-symmetric state, while the Hamiltonian is totally symmetric~\cite{Zanardi98}. Therefore, in a
combination of these limits coherent mixtures of these states are stable. However, the states are at different energies which gives 
rise to the exponential factors in Eq.~(\ref{eqgenevolt2}).

\section{Entanglement Dynamics}
\label{secentanglementdynamics}

We are now in a position to say that, provided our qubits are sufficiently close together 
such that $\delta\ll1$, for times $t>\gamma_0^{-1}$ contributions from eigenvectors with $l>3$ and their associated dynamics will 
have all but vanished, and to a good approximation (and a better approximation as time increases), 
our two-qubit state will be well described by Eq.~(\ref{eqgenevolt}). To quantify the resulting entanglement dynamics 
we use Wootters concurrence~\cite{wootters}, which ranges from 0 for separable states to 1 for maximally entangled states. It can be shown 
numerically that the concurrence of a state described by Eq.~(\ref{eqgenevolt}) depends only very weakly on the magnitude of $a_2$, and in the 
zero temperature and separation limit has completely vanishing dependence. We may therefore set $a_2=0$ to derive a simple concurrence expression, and from 
Eqs.~(\ref{alpha0}),~(\ref{alpha0alpha1}) and (\ref{eqgenevolt}) find
\begin{equation}
C=\mathrm{max}\Bigl[\frac{(R^2-1)(R^2+3)+(R^2-\Lambda)(3-R^2)\e^{\lambda_1 t}}{2(R^2+3)},0\Bigr],
\label{eqcontime}
\end{equation}
valid (with increasing accuracy) for timescales $t>\gamma_0^{-1}$. The legitimacy of setting $a_2=0$ 
will be demonstrated towards the end of this section, where 
comparisons with numerics using the full eigensystem of $\mathcal{L}$ are made. 

\begin{figure}
\begin{center}
\includegraphics[width=0.45\textwidth]{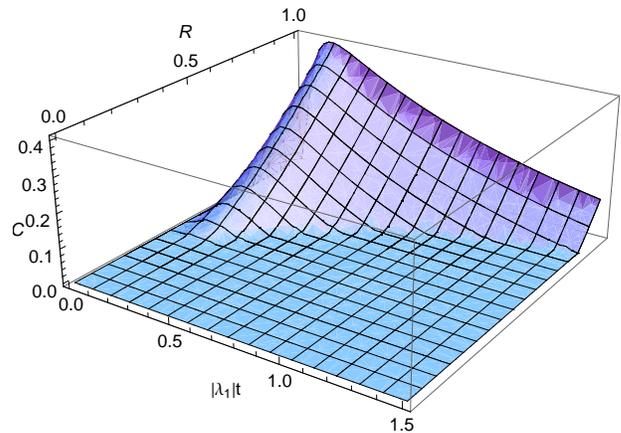}
\caption{(Color Online) Concurrence of the initially separable state $\ket{\uparrow \downarrow}$ in either $x$, $y$, or $z$ ($\Lambda=-1$) 
as a function of time (scaled by $\lambda_1$) and $R=\mathrm{tanh}(\Delta/2 k_B T_B)$ calculated from the full Liouvillian, though ignoring the Lamb-shift.}
\label{entangtimeR}
\end{center}
\end{figure}

Whether the bath is capable of inducing spin entanglement, and if so how  
it subsequently decays, is now clear. Firstly, no entanglement is  
generated unless 
\beq
\Lambda<\frac{5R^2-3}{3-R^2}, 
\label{eqnentanglementcondition}
\eeq
in agreement with Ref.~\cite{benatti05}. Secondly, provided this inequality is satisfied, we see from Eq.~(\ref{eqcontime}) that 
the induced entanglement will decay exponentially until  
$(R^2-\Lambda)(3-R^2)\e^{\lambda_1 t}=-(R^2-1)(R^2+3)$ is satisfied, after which time the entanglement is zero. This occurs at
\begin{equation}
t_c=\frac{1}{|\lambda_1|}\mathrm{ln}\left[\frac{(R^2-\Lambda)(R^2-3)}{(R^2+3)(R^2-1)}\right],
\label{eqtc}
\end{equation}
as demonstrated in Fig.~\ref{entangtimeR}. Note that Eqs.~(\ref{eqcontime}) and (\ref{eqtc}) are valid for a range of temperatures, 
however, in view of the inequality of Eq.~(\ref{eqnentanglementcondition}), we will focus on small temperatures since these maximise the amount 
and life-time of any induced entanglement. Interestingly, in the limit of vanishing temperature ($R\rightarrow1$), the entanglement 
reaches zero only asymptotically. Also, as the qubit separation $d \rightarrow 0$  
the level of entanglement becomes a function of the initial state only and $t_c\rightarrow \infty$, i.e. 
the steady-state becomes entangled~\cite{benatti05}. In general, $t_c$ varies as $(\delta\gamma_0)^{-1}$ for given $R$ and $\Lambda$, 
hence it can be lengthened by increasing the ratio 
$\Delta/T_B$, or by decreasing the separation $d$. 

We would naturally like to know which initially separable states result in the 
largest generated concurrence.  
For fixed temperature and spin separation 
the only parameter to consider is $\Lambda$, and from
inspection of Eq.~(\ref{eqcontime})  
we see that it should be minimised.  
This corresponds to an initial state that is as anti-symmetric as 
possible; hence, $\Lambda$ is minimised by anti-aligning the single spin 
Bloch vectors, giving $\Lambda=-1$ for pure states. Such a state corresponds to $\ket{\uparrow\downarrow}$ 
in $x$, $y$ or $z$. As the states become more mixed, the Bloch vectors 
decrease in length and $\Lambda\rightarrow0$. 
Interestingly, even a {\it maximally mixed state} ($\Lambda=0$) can become entangled provided $\Delta>2 \coth^{-1}(\sqrt{5/3}) k_B T_B$. 
In general, we expect an initially separable state to reach its maximum entanglement after a time 
$\sim\gamma_0^{-1}$, typical of the decay of eigenvectors $\vec{\alpha}_l$ for $l> 3$. Note that if we do not 
restrict our initial state to a separable state, $\Lambda$ is minimised by the Bell singlet, for which $\Lambda=-3$. As we have mentioned, 
in the limit that the qubit separation goes to zero the singlet is able to maintain its full entanglement for all times.

\begin{figure}
\begin{center}
\includegraphics[width=0.45\textwidth]{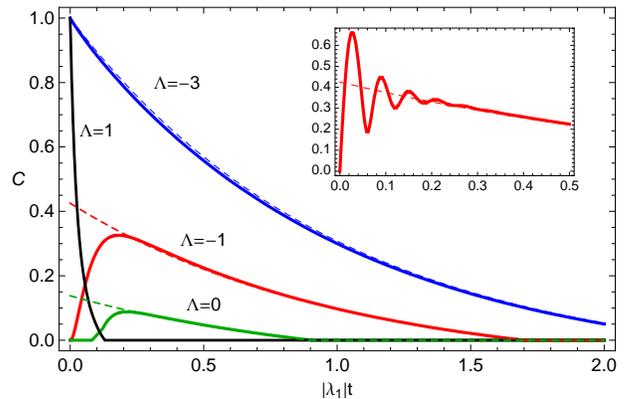}
\caption{(Color Online) Main: Concurrence as a function of (scaled) time calculated analytically (dashed lines, valid for $t>\gamma_0^{-1}$) and numerically (solid lines). We consider four initial states, the Bell 
singlet $(1/\sqrt{2})(\ket{\uparrow \downarrow}-\ket{\downarrow \uparrow})$ (blue, $\Lambda=-3$), the pure state $\ket{\uparrow \downarrow}$ 
(red, $\Lambda=-1$), the maximally mixed state $\rho=(1/4)I$ (green, $\Lambda=0$) and the Bell state 
$(1/\sqrt{2})(\ket{\uparrow \downarrow}+\ket{\downarrow \uparrow})$ (black, $\Lambda=1$). Parameters: $\delta=0.05$, 
$R=0.9$. Inset: Behaviour of the initial state $\ket{\uparrow \downarrow}$ where, in the numerical calculations, the Lamb-shift and exchange interactions
have been included at an arbitrarily chosen strength $B=\xi=1/(2|\lambda_1|)$.}
\label{figentanggraph}
\end{center}
\end{figure}

To illustrate these points, in the main part of Fig.~(\ref{figentanggraph}) we plot the time evolution of the concurrence 
for various initial states, calculated both from Eq.~(\ref{eqcontime}) (dashed lines) and numerically using the full Liouvillian 
(solid lines), though here ignoring the Lamb-shift. As claimed, on timescales $>\gamma_0^{-1}\sim\delta$ in the scaled time units, 
the entanglement dynamics is well approximated by the analytic form. Note also that neglecting the eigenvectors $\vec{\alpha}_2$ 
and $\vec{\alpha}_3$ has  
had no discernible effect on the accuracy of Eq.~(\ref{eqcontime}) on the timescales 
it is expected to be valid. For the initially separable states ($\Lambda=-1,0$), we clearly 
see that the entanglement reaches its maximum on a timescale $>\gamma_0^{-1}\sim\delta$, decaying subsequently at a rate $\sim\delta\gamma_0$. 
For the Bell singlet ($\Lambda=-3$), the analytic approximation becomes almost exact since 
this state is simply a linear combination of the vectors $\vec{\alpha}_0$ and $\vec{\alpha}_1$ in Eqs.~(\ref{alpha0}) and~(\ref{alpha0alpha1}). 
Also shown is the behaviour 
of the Bell state $(1/\sqrt{2})(\ketz{\uparrow \downarrow}+\ketz{\downarrow \uparrow})$, for which $\Lambda=1$. Unlike the singlet, 
all other Bell states have the maximum possible value of $\Lambda$, and as such lose their entanglement rapidly. 

We could consider the maximally mixed state ($\Lambda=0$) as being the infinite 
temperature thermal state since it represents a state for which thermal fluctuations completely overcome any external fields. We see that
as the bath ``cools'' this state towards the thermal state at  
$T_B$, it does so via entangled states. 
Of course, 
after a time $t_c$ 
the state of the qubits becomes separable once more, and will eventually reach the thermal state at $T_B$. There is  
in fact a well 
defined condition for $T_B$ and (initially prepared) qubit temperature $T_Q$ such that the bath has the ability to entangle 
the qubits. In the limit of small bath and qubit temperature this condition becomes
$\theta_B-\theta_Q>\frac{1}{2}\mathrm{ln}3$,
where $\theta_B=(\Delta/2 k_b T_B)$ and $\theta_Q=(\Delta/2 k_b T_Q)$. 

\section{Entanglement generated through the Lamb-shift}\label{sec:lamb}

It is important to consider the role of the Lamb-shift Hamiltonian $H_{LS}$, thus far ignored. 
Within the limits of our derivation, namely that $\Delta$ is large and hence rotations about $x$ are so rapid that the $y$- and $z$- directions are 
effectively equivalent, its form is determined by symmetry. 
With this form it commutes with our eigenvectors of most interest, $\vec{\alpha}_0$ and $\vec{\alpha}_1$. It can also be shown that the 
eigenvectors $\vec{\alpha}_2$ and $\vec{\alpha}_3$ are eigenoperators of $H_{LS}$, with eigenvalues $2(A+B)$ and $-2(A+B)$, respectively. 
Therefore, the Lamb-shift Hamiltonian can influence only the eigenvectors with $l>3$, 
the imaginary parts of their eigenvalues, and the imaginary parts of $\lambda_2$ and $\lambda_3$. 
Hence, despite the fact that $H_{LS}$ can entangle our spins, its effect 
is restricted to a timescale of order $\gamma_0^{-1}$ (after which the other eigenvectors have decayed) regardless of its amplitude, and it will 
therefore have no effect on the long-time entanglement dynamics  
or the analytic expressions we have derived. 

In fact, we can further account for the effect of a direct spin exchange interaction simply by adding a term of the form $H_E=\xi\vec{\sigma}^1\cdot\vec{\sigma}^2$ into  
$H_S$ in Eq.~(\ref{eqrhodot}), provided that the evolution it generates occurs on timescales much slower than the bath correlation time. This 
procedure is valid in the regime of $\Delta\gg\xi$, such that $\Delta$ sets the relevant frequency scale for the system-bath interaction.
In this case, exactly the same 
conclusions hold as for the Lamb-shift term since $\vec{\alpha_0}$ and $\vec{\alpha_1}$ again commute with this form of interaction, and 
$\vec{\alpha}_2$ and $\vec{\alpha}_3$ are also eigenoperators of $H_E$. Its influence 
will thus similarly be restricted to short timescales $\sim\gamma_0^{-1}$. This point is illustrated in the inset of Fig.~(\ref{figentanggraph}) 
where we plot the analytically and numerically calculated concurrence of the initial state $\ket{\uparrow \downarrow}$ in 
$x$, $y$ or $z$, including both $H_{LS}$ and $H_E$ with arbitrarily chosen strengths. 
As expected, their impact is seen only on a timescale $\sim\gamma_0^{-1}$, much shorter than that over which the dissipatively 
induced entanglement survives. 


\section{Experimental Realisation}

While spin-boson models apply commonly in solids, more controlled realisations in ion traps have recently been 
proposed~\cite{porras08}. We extend to the two-spin-boson model as follows: we consider a linear ion trap
with the internal levels of a single ion representing each spin, 
coupled to the collective motion of $N$ ions providing a (finite) bosonic bath. 
The 
two-spin-boson Hamiltonian is generated by simultaneously 
addressing two ions with traveling waves which, in a linear ion trap, gives rise to an Ohmic spectral density $J(\omega)=(\alpha/2)\omega$. 
The static field $\Delta$ is set by the laser-ion Rabi frequency. The strength of the system-bath interaction can be adjusted by varying various 
experimental parameters such as the laser wavelength and ion mass. We work in the weak coupling regime and therefore set $\alpha=0.1$~\cite{leggett87}.
Addressing the ions in the way described also induces an effective Ising interaction between the 
two ions due to the polaron representation~\cite{porras08}, which nevertheless disappears for a finite lattice 
with Ohmic spectral density.

Owing to the finite size of the bath, the system evolves as if it were coupled to a 
continuum only for short times, after which a quantum revival is seen. For example, for $N=100$ this revival occurs at 
approximately $2 \pi /\omega_t$, where $\omega_t$ is the trapping frequency~\cite{porras08}. Hence, to observe both the generation and subsequent decay 
of the dissipatively-induced entanglement we describe, this period must be larger than both $\gamma_0^{-1}$ and $(\delta \gamma_0)^{-1}$. Since 
both $\gamma_0$ and $\delta$ depend on $\Delta$, this corresponds to a careful choice of $\Delta$.
With $\Delta=25\omega_t$ (set by the Rabi frequency), we find a revival time $\sim50\gamma_0^{-1}$.
Assuming the wavelength associated with $\Delta$ to be approximately $N/(\Delta/\omega_t)$ in units of the ion spacing, we choose to address two neighbouring ions to give 
$\delta=(\Delta/\omega_tN)^2/2D \approx 0.03$.

Furthermore, the temperature of the bath must be low enough such that the inequality of Eq.~(\ref{eqnentanglementcondition}) is 
satisfied, allowing a finite level of entanglement to be generated.
By requiring $R=1/2$  we find that the bath must have a temperature in the mK range for typical trapping frequencies of MHz. 
We conclude that after a timescale $\sim\gamma_0^{-1}=(25 \pi \alpha \omega_t)^{-1}$  
a concurrence of $C\approx0.15$ should be generated from the initial state $\ket{\uparrow \downarrow}$ in $x$, $y$ or $z$. 
It will subsequently decay by at least a factor of $\mathrm{e}^{-1}$ before the dynamics associated with the finite size of the bath becomes significant. 

\section{Summary}

We have shown that the mechanisms normally associated with dissipative processes can lead to long-lived 
entanglement in {\it non-interacting, spatially separated} two-qubit systems.  
We have highlighted two important timescales. The first, shorter timescale $\gamma_0^{-1}$ is that with which we 
expect a single qubit to dephase. When a second qubit is introduced close to the first, we find dissipatively-induced 
entanglement is generated on this short timescale, and further that there is a second 
larger timescale $(\delta\gamma_0)^{-1}$ on which the induced entanglement decays. Importantly, 
the influence of both the bath-induced Lamb-shift or a direct spin exchange interaction is still restricted to the original shorter timescale. 
Hence, the presence of a second qubit within the bath induces coherences in the overall system state that can persist on timescales 
far larger than either the corresponding single qubit decoherence time, or timescales associated with the influence of direct exchange or the Lamb-shift. 

\subsection{Acknowledgements}

We thank H. T. Ng for useful discussions. SB thanks the Royal Society and Wolfson Foundation. DPSM, AN and SB are supported by the {\sc epsrc}.

\end{document}